%
%
%
%
%

\input harvmac
\input epsf
 
\parindent=0pt
\Title{SHEP 97-08}{\vbox{\centerline{C Function Representation}
\vskip2pt\centerline{of the}
\vskip2pt\centerline{Local Potential Approximation}}}

\centerline{\bf Jacek Generowicz, Chris Harvey-Fros and Tim R. Morris}
\vskip .12in plus .02in
\centerline{\it 
Department of Physics, University of Southampton,}
\centerline{\it Highfield, Southampton SO17 1BJ, UK}
\vskip .7in plus .35in

\centerline{\bf Abstract}
\smallskip 
Within the Local Potential Approximation to Wilson's, or Polchinski's,
exact renormalization group, and for general spacetime dimension,
we construct a function, $c$, of the coupling constants;
it has the property that (for unitary theories)
it decreases monotonically along flows, and is stationary
only at fixed points ---where it `counts degrees of freedom', {\it i.e.}
is extensive, counting one for each Gaussian scalar. Furthermore,
by choosing restrictions to some sub-manifold of coupling constant space, 
we arrive at a very promising variational approximation method.

\vskip -1.5cm
\Date{\vbox{
{hep-th/9705088}
\vskip2pt{May, 1997.}
}
}

\def\ins#1#2#3{\hskip #1cm \hbox{#3}\hskip #2cm}
\def\etc{{\it etc.}\ }
\def\ie{{\it i.e.}\ }
\def\eg{{\it e.g.}\ }
\def\cf{{\it cf.}\ }
\def\viz{{\it viz.}\ }
\def\aka{{\it a.k.a.}\ }

\def\phi{\varphi}

\def\E{{\cal E}}
\def\F{{\cal F}}
\def\G{{\cal G}}
\def\hG{{\hat{\cal G}}}
\def\M{{\cal M}}

\parindent=15pt

\noindent{\bf 1. A ${\bf C}$ Function.}

Following Zamalodchikov's celebrated 
$c$-theorem\ref\zam{A.B. Zamolodchikov, JETP Lett. 43 (1986) 730.}\
for two dimensional quantum field theory, a number of groups have sought
to generalise these ideas to higher 
dimensions\ref\zamfan{J.L. Cardy, Phys. Lett. B215 (1988) 749;\
H. Osborn, Phys. Lett. B222 (1989) 97;\
I. Jack and H. Osborn,
Nucl. Phys. B343 (1990) 647;\ 
A. Cappelli {\it et al}, Nucl. Phys. B352 (1991) 616,
B376 (1992) 510;\  D. Freedman {\it et al},
Mod. Phys. Lett. A6 (1991) 531;\
G.M. Shore, Phys. Lett. B253 (1991) 380, B256 (1991) 407;\ 
A.H. Castro Neto and E. Fradkin, Nucl. Phys. B400 (1993) 525;\ 
P. Haagensen {\it et al}, Phys. Lett. B323 (1994) 330;\ 
X. Vilas\'\i s-Cardona, Nucl. Phys. B435 (1995) 635;\ 
A.C. Petkou, Phys. Lett. B359 (1995) 101;\ 
V. Periwal, Mod. Phys. Lett. A10 (1995) 1543;\ 
J. Gaite and D. O'Connor, Phys. Rev. D54 (1996) 5163;\ 
J. Gaite, C96-08-26.1, hep-th/9610040;\
F. Bastianelli, Phys. Lett. B369 (1996) 249;\
J. Comellas and J.I. Latorre, UB-ECM-PF 96/1, hep-th/9602123;\
R.C. Myers and V. Periwal, PUPT-1567, hep-th/9611132;\
J.I. Latorre and H. Osborn, DAMTP-97-1, hep-th/9703196.}. The
motivation behind this, is not only to demonstrate irreversibility
of renormalization group flows, in some precise sense and under certain
general conditions, and thus prove that exotic flows such as limit cycles,
chaos, \etc are missing in these cases, but perhaps more importantly to
provide an explicit, and useful, geometric framework for the space of
quantum field theories.
In ref.\zam, Zamalodchikov established three important properties for his
$c$ function,
\medskip

\item{1.} There exists a function $c(g)\ge0$ of such a nature that
\eqn\cfl{ {d\over dt} c\equiv \beta^i(g){\partial\over\partial g^i} c(g)
\le 0\quad,}
where the $g^i$ form an (infinite) set of dimensionless parameters,
\aka the coupling constants, and the beta functions 
are defined as $\beta^i=dg^i/dt$.
The equality in \cfl\ is reached only at fixed points 
of the renormalization group $g(t)=g_*$.
\item{2.} $c(g)$ is stationary at 
fixed points;\foot{We will not here need to restrict to 
critical fixed points\ref\kogwil{K. Wilson and J. Kogut, Phys. Rep. 
12C (1974) 75.}.}\ 
\ie $\beta^i(g)=0$ for all i, implies $\partial c/\partial g^i=0$.
\item{3.} The value of $c(g)$ at the fixed point $g_*$ is the same as
the corresponding (Virasoro algebra) central charge\ref\cft{A.A. 
Belavin, A.M. Polyakov and A.B. Zamolodchikov, Nucl. Phys.
B241 (1984) 333.}. (This property thus
only makes sense in two dimensions.)
\medskip

Within the Local Potential Approximation\ref\LPA{J.F. Nicoll,
T.S. Chang and H.E. Stanley, Phys. Rev. A13 (1976) 1251.}\nref\zum{G.
 Zumbach, Nucl. Phys. B413 (1994) 754, Phys. Lett. A190 (1994) 
225.}\nref\ball{R.D. Ball {\it et al}, Phys. Lett. B347 (1995) 
80.}\nref\truncm{T.R. Morris, Nucl. Phys. B458[FS] (1996)
 477.}--\ref\revii{T.R. Morris,  in {\it RG96}, SHEP 96-25,
 hep-th/9610012.}(LPA) to Wilson's\kogwil\ or
Polchinksi's\ref\pol{J. Polchinski, Nucl. Phys. B231 (1984) 269.}\ 
exact renormalization group (which we describe
below), we display a
$c$-function which has the first two properties in any dimension $D$.
This follows by virtue of the
property that the $\beta$-functions 
are the `gradient flows' of $c$ with respect
to a positive definite metric. This property 
in turn follows straightforwardly
from an observation by Zumbach\zum. 
At fixed points,  
our $c$ function is extensive, \viz additive in
mutually non-interacting degrees of freedom,
as is also true of the Virasoro central charge (\cf property 3).
Furthermore, we may normalise so that
our  $c$ counts one for each Gaussian
scalar and zero for each infinitely massive scalar (corresponding to a
High Temperature fixed point), as does the central charge in two dimensions.
It is probably not possible within the Local Potential Approximation, 
to establish a more concrete link to Zamalodchikov's $c$. 
Of course, it would be very 
interesting to understand if these observations generalise
to higher orders in the derivative expansion\ref\twod{T.R. Morris, 
Phys. Lett. B345 (1995) 139.}\revii\ 
(which likely would allow a direct comparison with Zamalodchikov's
$c$), or indeed generalise to an exact expression along the present lines.
  
In the second part of this letter, we compute exactly within the LPA,
some simple illustrative examples.
In the third and final part, 
we point out that this geometrical structure
for the LPA leads naturally to a variational approximation scheme,
by restricting the flow to some finite dimensional
submanifold of coupling constant space, of our choosing.
Approximate continuum limits (\ie fixed point behaviour), 
follow from stationarizing with respect to this finite parameter set. 
We briefly investigate the efficacy of the method, considering, in 
particular, the simplest polynomial approximation to the non-perturbative
fixed point potential for a single scalar field in three dimensions.
The resulting form, and critical exponents, lie exceedingly close to
the exact\foot{within LPA!}\ answers.

We work with $N$ Bose fields $\phi_a$ in $D$ Euclidean dimensions. We have
checked that the constructions can be generalised 
to Fermi fields, but for simplicity we omit this.
In a condensed notation\ref\erg{T.R. Morris, Int. J. Mod. Phys. A9 (1994) 2411.}\ref\largen{
M. D'Attanasio and T.R. Morris, SHEP 97-03, hep-th/9704094.},
Polchinski's form of the Wilson renormalization group is given
by\pol\erg\ref\zin{J. Zinn-Justin, 
            ``Quantum Field Theory and Critical Phenomena'' (1993)
             Clarendon Press, Oxford.}\
\eqn\Pol{
{\partial S_\Lambda \over\partial\Lambda}={1\over2}\,
 {\delta S_\Lambda\over\delta\phi_a}\cdot
{\partial\Delta_{UV}\over\partial\Lambda}\cdot
{\delta S_\Lambda\over\delta\phi_a}
-{1\over2}\, {\rm tr}\, 
{\partial\Delta_{UV}\over\partial\Lambda}\cdot
{\delta^2S_\Lambda\over\delta\phi_a\delta\phi_a}
\quad.}
Here $S_\Lambda[\phi]$ is the interaction part 
 of a Wilsonian effective action
$S^{eff}_\Lambda=
\half\phi_a.\Delta_{UV}^{-1}.\phi_a +S_\Lambda$, and
$\Delta_{UV}=C_{UV}(q^2/\Lambda^2)/q^2$
 is a massless propagator whose momentum
$q$ is cutoff by an effective ultra-violet cutoff $\Lambda$.
We require that the cutoff function $C_{UV}$ is analytic at $q=0$ and
that $C_{UV}(0)=1$ (so that physics is unchanged at scales much less than
$\Lambda$), and $C_{UV}\to0$ as $q\to\infty$ sufficiently rapidly that 
all momentum integrals in \Pol\ are well-defined (\ie well regulated).
Wilson's flow equation\kogwil\
is identical to \Pol, after the transformation\ref\deriv{T.R. Morris, 
Phys. Lett. B329 (1994) 241.}\revii\
\eqn\PtoW{\phi\mapsto\sqrt{C_{UV}}\,\phi, \ins11{and identification}
{\cal H}\equiv- S_\Lambda\quad.} 

In the Local Potential Approximation, we restrict the interactions to that
of a general potential:
$S_\Lambda=\int\!d^D\!x\, V(\phi,\Lambda)$, and discard from the right hand
side of \Pol\ all terms that correspond to higher derivative corrections
(on expansion of $C_{UV}$ where necessary). Thus,
$$\eqalign{{\partial\over\partial\Lambda} V(\phi,\Lambda) &=
{\alpha\over\Lambda^3} \left({\partial V\over\partial\phi_a}
\right)^2 -\gamma\Lambda^{D-3}
 {\partial^2V\over\partial\phi_a\partial\phi_a}\quad,
 \cr
\ins01{where}\alpha &=-C'_{UV}(0)\ins{.5}{.5}{and}
\gamma= -\int\!\!{d^D{\tilde q}\over(2\pi)^D}\, C'_{UV}({\tilde q}^2) 
\quad.\cr}$$
We assume that the dimensionless coefficients $\alpha$ and $\gamma$
are positive; this is the case, for example, if $C_{UV}$ is strictly
monotonically decreasing. To express all quantities in a form
appropriate for the continuum limit (\ie approach to fixed points),
 we change to dimensionless\foot{in general using $\Lambda$
raised to the power of their scaling dimensions,
however in LPA there is no anomalous dimension for $\phi$.
This fact also allowed our scaling-dimensionless choice for $C_{UV}$ \deriv.}\
 variables 
$\phi\mapsto \phi\Lambda^{D/2-1}\sqrt{\gamma}$,
$V\mapsto V\Lambda^D\gamma/\alpha$, and $t=\ln(\mu/\Lambda)$
(where $\mu$ is some arbitrary physical mass scale), and 
by thus also absorbing the coefficients $\alpha$ and $\gamma$, obtain
the LPA of Polchinski's equation
\eqn\lpa{
{\partial\over\partial t} V(\phi,t) +{1\over2}(D-2)\phi_a{\partial V\over
\partial\phi_a} - D V = 
 {\partial^2V\over\partial\phi_a\partial\phi_a}
-\left({\partial V\over\partial\phi_a}\right)^2 \quad,}
in a manifestly `scheme' independent form\ball. 
Since the LPA corresponds to discarding all momentum dependent
terms from \Pol,  it is immediate to realise from \PtoW\ that the 
above is also the LPA  for Wilson's equation. 

If in place of $V$, we introduce a Gibbsian-like measure 
$\rho(\phi,t) =\exp\left\{-V(\phi,t)\right\}$,
and $G(\phi)=\exp\left\{-{1\over4}(D-2)\phi_a^2\right\}$,
and define the functional
\eqn\defF{ \F[\rho] =a^N\!\!\!\int\!\! d^N\!\phi\ G\left\{
{1\over2}\left({\partial\rho\over\partial\phi_a}\right)^2
+{D\over4}\rho^2\left(1-2\ln\rho\right)\right\} \quad,}
then \lpa\ may be rewritten manifestly as a `gradient flow' \zum:
\eqn\zums{ a^NG\,{\partial\rho\over\partial t}
=-{\delta\F\over\delta\rho}   \quad.} 
In \defF\ and \zums, $a>0$ is a normalisation factor for the $\phi$ measure,
to be determined later. [Alternatively, changing  variables
$\phi\mapsto\phi/a$, the factors $a^N$ can be absorbed and appear 
instead as factors of $a^2$
in front of the $(\partial/\partial\phi)^2$ terms in \lpa\ and \defF.]
Substituting the fixed point equation $\delta\F/\delta\rho=0$ back into
\defF, we see that fixed points $\rho(\phi,t)=\rho_*(\phi)$ satisfy,
\eqn\fF{ \F[\rho_*]={D\over4}\,a^N\!\!\!\int\!\! d^N\!\phi\,G\rho^2_*\quad.}

Now let $g^i(t)$ be a complete set of parameters (\aka coupling constants)
for $V$. Away from fixed points, these will
 be infinite in number. Asymptotically
approaching a particular fixed point, it is possible to identify a finite
number of parameters (namely the relevant and marginal couplings) 
and express all the rest
(the `irrelevant'
 parameters) in terms of them. (This `reduction of parameters'
corresponds to the 
continuum limit defined around this fixed point, see 
later and \eg \erg\ref\eqs{T.R. Morris, SHEP 96-36, 
hep-th/9612117, to be published in Nucl. Phys. B};
universality arises because perturbations in the irrelevant
parameters about these values, exponentially decay away as $t\to\infty$.)
Following ref.\zam, we define the operators
\eqn\op{ \Phi_i(g)=\partial_i V(\phi,g)\quad,}
where we have written $\partial_i\equiv{\partial/\partial g^i}$.
We define the metric
\eqn\defG{\G_{ij}(g)=a^N\!\!\!\int\!\! d^N\!\phi\,
G\rho^2\,\Phi_i\Phi_j\quad.}
It is positive definite providing $V$ is real, which we assume
 (corresponding to a 
unitary theory in Minkowski space).
Multiplying \zums\ by $\partial_i\rho=-\rho\Phi_i$, we thus obtain
\eqn\bfl{ \G_{ij}\beta^j=-\partial_i\F(g)\quad.}

We define the $c(g)$ function through
\eqn\defc{\F= D A^{c}/4\quad,}
where $A>1$ is a normalisation factor to be determined.
Redefining the metric by a positive factor,
$\G_{ij}= (\F\ln A)\, \hG_{ij}$,
 we obtain from \bfl:
\eqn\bcfl{\partial_i c(g)=-\hG_{ij}\,\beta^j(g)\quad.}

From \bcfl\ and the positive-definiteness of $\hG_{ij}$, 
we see that \cfl, and the properties 1 and 2 hold.
Now suppose that the fields 
form two mutually non-interacting
sets. Let us write $\phi_a\equiv\phi_a^{(1)}$ when the field belongs
to the first set, and $\phi_a\equiv\phi_a^{(2)}$ when it belongs
to the second set. Similarly, the couplings $g$ split into two sets
$g^{(1)}$ and $g^{(2)}$. 
The potential $V(\phi,t)$ can be written
$V(\phi,t)=V^{(1)}(\phi^{(1)},t)+V^{(2)}(\phi^{(2)},t)$.
Thus $\rho$ factorizes: $\rho =\rho^{(1)}\rho^{(2)}$, 
and from \fF, at fixed points
we have $\F[\rho_*]={4\over D}\F[\rho_*^{(1)}]\F[\rho_*^{(2)}]$.
Therefore from \defc\ we have, at fixed points, that our $c$ is
extensive: 
\eqn\ext{c(g_*)=c(g^{(1)}_*)+c(g^{(2)}_*)\quad.}

At the Gaussian fixed point $V_*=0$, the Virasoro central charge
counts one degree of freedom per scalar, \ie 
is here equal to $N$ \cft. At the High Temperature fixed point 
$V_*=V^{HT}_*={1\over2}\phi_a^2-N/D$,
the Virasoro central charge vanishes,
because  $V^{HT}_*$ corresponds to a non-critical fixed point, \ie
infinitely massive\foot{in units of $\Lambda$}\  fields and
thus no propagating degrees of freedom.
 This interpretation of $V^{HT}_*$ follows directly from the
Legendre transform relation between the Polchinski and Legendre
effective potentials\largen\erg,  or by the explicit flow derived below
(see also ref.\ref\com{J. Comellas and A. Travesset, UB-ECM-PF 96/21, hep-th/9701028.}).

The normalisation factors $A$ and $a$ may be uniquely determined
by requiring that our $c$ agree with this counting in any dimension $D$. 
Substituting $c=0$ in \defc\ and $\rho_*=\exp-V^{HT}_*$ in \fF, we
obtain 
\eqn\defa{a=\e{-2/D}\,\sqrt{{D+2\over4\pi}}\quad.}
For the Gaussian fixed point we substitute $c=N$ in \defc\ and 
$\rho_*=1$ in \fF, and thus
\eqn\defA{A= \e{-2/D}\,\sqrt{{D+2\over D-2}}\quad.}
Note that $A>1$, as required, for all $D\ge2$.
The precise values for $a$ and $A$ may be expected to change
beyond the LPA. 

Perturbing the couplings to first order
about a fixed point: $g^i(t)=g_*^i+\epsilon v^i\e{\lambda t}$, we obtain
from \bcfl\ the eigenvalue equation
$$ \partial_i\partial_j c(g_*)\,v^j
=-\lambda\,\hG_{ij}(g_*)\,v^j\quad,$$ 
determining the eigenoperators $\Phi=v^i\Phi_i$ and their scaling dimensions
$D-\lambda$. Equivalently, from \bfl,
\eqn\ev{ \partial_i\partial_j \F(g_*)\,v^j=-\lambda\,\G_{ij}(g_*)\,v^j\quad.}

\medskip
\noindent{\bf 2. Examples.}

Consider the simple example of the Gaussian fixed
point perturbed by the mass operator, for a single scalar field.
Thus we set $V(\phi,t)=\half\sigma(t)\phi^2
+\E(t)$. The $\beta$ functions for $\sigma$ 
and $\E$ follow easily from \lpa:
\eqn\twobe{
\eqalign{ {\partial\over\partial t} \sigma &= 2\sigma (1-\sigma) \cr
{\partial\over\partial t} \E &= D\E+\sigma \quad.\cr}}
The general solutions are \eqna\gausol\
$$\eqalignno{ \sigma(t) &={1\over 1+r\, \e{-2t}} &\gausol a\cr
\E(t) &=\E_0\,\e{Dt}-{1\over2} \int^1_0\!\!\!\! du \, 
{u^{D/2-1}\over 1+ur\,\e{-2t}}\quad, &\gausol b\cr}$$
where $r$ and $\E_0$ are integration constants. Considering
$t\to-\infty$, one sees that this solution indeed emanates from the Gaussian
fixed point, while a
change of variables back to dimensionful (\ie physical) variables 
shows that
$r\sim \mu^2/m^2$, where $m$ is the
mass of the still-Gaussian  scalar field, while $\E_0\mu^D$ is an added
vacuum energy term. We normalised the special solution in \gausol{b}\
so that with $\E_0=0$, \gausol{}\ tends to $V^{HT}_*$ as $t\to\infty$.

Since $\rho$ is Gaussian, \defF\ is readily determined;
\eqn\twoF{\F={a\over2}\,
\e{-2\E}\left\{2D\E+{6D\sigma +4\sigma^2+D^2-2D\over D-2+4\sigma}\right\}
\sqrt{{\pi\over D-2+4\sigma}} \quad.}
Combining this, \defa, \defA\ and \defc, yields $c(t)$ and one
verifies that when $\E_0=0$, $c(t)$ flows from 1 to 0 as $t$ runs from
$-\infty$ to $+\infty$. Note that if $\E_0\ne0$, then $c(t)\to\mp\infty$
as $t\to\infty$, depending on the sign of $\E_0$.
This seems in contradiction with the idea that $c$ counts
degrees of freedom, since evidently the vacuum energy should not figure
{\it per se} in this counting. However, one must recall from \ext, that 
$c$ is extensive (\ie `counts') only at fixed points, and with $\E_0\ne0$
the system never reaches another fixed point as $t\to\infty$.

From the Gaussian form of $G$, we recognize that at both the Gaussian and High
Temperature fixed points, the metric \defG\ (and thus $\hG_{ij}$)
is diagonalized by choosing the operators $\Phi_i$ to be products of
Hermite polynomials $H_n$ in the $\phi_a$. Since these also turn out
to diagonalize $\partial_i\partial_j\F(g_*)$, they are the eigenperturbations,
and the corresponding eigenvalues follow straightforwardly. Choosing 
the Gaussian fixed point
$\rho_*=1$ for example, and
again specializing to the case of
one scalar field for simplicity, we thus take 
$\Phi_n=H_n\left({\phi\over2}\sqrt{D-2}\right)$, 
$n=0,1,\cdots$. The metric has non-zero components\ref\GR{See
 \eg I.S. Gradshteyn and I.M. Ryzhik,
 ``Table of integrals, series, and products'' (1980) Academic Press.}\
$\G_{nn}=2^{n+1} n!a\sqrt{{\pi\over D-2}}$ . From \op\ and \defF, we
obtain 
$$ \partial_i\partial_j\F\Big|_{\rho=1}=-D\,\G_{ij}+a\!\int\!\! d\phi\ G\,
 {\partial\Phi_i\over\partial\phi}{\partial\Phi_j\over\partial\phi}\quad.$$
Thus, using\foot{prime being differentiation with respect to the argument}\
 $H'_n=2n H_{n-1}$, we have that $\partial_i\partial_j\F$ is also diagonal, and
we recover the expected Gaussian spectrum of eigenvalues 
$\lambda =-\partial_n\partial_n\F/ \G_{nn}=D+\half(2-D)n$.

We may also construct perturbatively, the
continuum limit  about the Gaussian fixed point, and
$c$ and $\G_{ij}$, to any desired order in the relevant
and marginal couplings. Thus for example in $D=4$ dimensions, the
massless continuum limit is constructed by solving the $\beta$ functions
for $g^2$, $g^6$, $g^8$, \etc, iteratively in terms of
a power series in $g^4(t)$ (in direct
generalisation of the case for ``$\lambda(t)$'' and ``$\gamma_1(t)$''
given in ref.\erg), reducing \bcfl\  to the one beta function
$\beta^4(g^4)=-\partial_4c(g^4)/\hG_{44}$.

\medskip
\noindent{\bf 3. Variational Approximations.}

The gradient flow form \zums\ of the LPA, suggests the possibility
of approximating $\rho$ by a variational ansatz, 
that is setting $\rho={\tilde\rho}(\phi;g_1,\cdots,g_M)$, $M<\infty$,
where ${\tilde\rho}$ is some finitely parametrized 
set of functions
of our choosing. Interpreting the functional derivative in \zums\
to include only variations in this restricted set, we arrive again at
\bfl\ [or \bcfl] where however, {\sl here and from now on}, 
indices run only from $1$ to $M$.
Thus geometrically, we restrict the flows
to the sub-manifold $\M$ parametrized by $g_1,\cdots,g_M$.  

Approximations to the fixed points of \lpa\ follow, by \bfl,
from solutions to
 the variational conditions $\partial_i\F=0$,\foot{since the 
(restriction of the) metric $\G_{ij}$ is still positive definite}
and correspond geometrically to those points $g={\tilde g}_*$ 
on $\M$ where $T_g\M$ (the tangent space to
$\M$ at $g$) is perpendicular
to the exact\foot{\ie with $M=\infty$, or from \lpa\ }\ flows.
From this, it is intuitively clear that 
good results can be expected generically, \eg for the shape of $\rho$, if
$\M$ passes sufficiently `close'
to the true fixed points, and good results will be obtained
for the scaling dimensions and shape, of any such
eigenoperators that are almost parallel with
 $T_{{\tilde g}_*}\M$. We can  expect decent
approximations for the full flows,
from say, one fixed point to another -- under similar conditions.

It is clear that it is possible
to be unlucky and find in this way `spurious fixed points' that do
not well approximate the exact solutions, but these will not be
relatively stable under changing or improving
the ansatz manifold $\M$, \eg by enlarging its dimension $M$.
In fact as $\M$ improves,
such spurious fixed points (if indeed there are any)
will disappear entirely.
Equally, it is clear that numerical results can (at least in principle)
be obtained to any desired degree of accuracy.

These characteristics are
 in marked contrast to the general situation for {\sl truncations}
(of the renormalization group) to a finite set of 
operators\ref\trunc{T.R. Morris, Phys. Lett. B334 (1994) 355.}\revii, 
where spurious fixed points multiply
at higher orders, and numerical results cease to converge.

To test the practicality of this variational
 method, we tried the following very simple even
polynomial ans\"atze. The simplest two cases, $V(\phi,t)=\E(t)$ and
$V(\phi)=\E(t)+\half\sigma(t) \phi^2$, both correspond to exact reductions
of the full flow equations (\aka $\beta$ functions) from \lpa,
and are given by \twobe. Therefore it is immediate to realise that
 we obtain only ---and exactly---
 the Gaussian and, when $\sigma$ is included, 
High Temperature fixed points.  Similarly, we obtain the
eigenvalue $\lambda=D$ for the unit operator $\Phi_\E=
\partial_\E V=1$ at both fixed points, and $\lambda=2$ (-2) for the second 
eigenvalue at the Gaussian (High Temperature)
fixed point when $\sigma$ is included. This may be checked directly
through \twoF, \ev, and \defG. 

In fact it is straightforward
to confirm that $\partial_\E V=1$, $\lambda=D$, is always a solution 
in any approximation that includes $\E$ as a parameter. Equally, from
our previous exact analysis of the Gaussian and High Temperature fixed
points, it is clear that a general polynomial ansatz of order
$\phi^{n}$, $n>0$, with unconstrained coefficients, will find exactly these
fixed points and exactly the first $n$ eigenvalues. 

The next-simplest even polynomial is $V(\phi,t)=\E(t)+\half\sigma(t)\phi^2
+s(t)\phi^4$. Since the case $s=0$ has already been analysed, we 
fix $s>0$. We set $D=3$. We take advantage of coordinate
invariance to write $s=\left(g^4\right)^2$ and $\sigma=g^4g^2-1/4$,
and change variables $\phi=x/\sqrt{g^4}$, so that $\F$ and its
derivatives may be expressed in closed form in terms of the integrals
$I_n(g^2)=\int^\infty_{-\infty}\!\!\!dx\, x^n\,\e{-g^2x^2-2x^4}$, $n=0,2$.
In this way, on the basis of analytic 
(for large $\pm g^2$) and numeric estimates, we establish 
that there is only one non-trivial
solution of the variational equations $\partial_i\F=0$, $i=\E,2,4$. It
corresponds to $s=.00772624$, $\sigma=-.13488$ and $\E=.054794$. 
In  fig.1, we plot
the resulting form for $\rho_*$ and compare it to the one exact non-trivial
fixed point solution\foot{Solved by shooting as described in
refs.\trunc\ball.}\ from \lpa\ (which itself is an approximation to
the Ising model fixed point in three dimensions). Solving \ev, we obtain,
apart from $\lambda=3$, $\nu=1/\lambda=.6347$ from the positive eigenvalue,
and $\omega=-\lambda=.6093$ from the remaining eigenvalue. These are
2\% and 8\% off the exact values from \lpa, namely $\nu=.6496$ and 
$\omega=.6557$, respectively \ball. 
Overall, we find these results, for this
simplest possible variational approximation to the non-trivial
fixed point (and/or $\omega$), truly impressive.

Since the fixed point behaviour of \lpa\ for a {\sl single} field
---or more generally--- a single invariant, is 
straightforward to solve directly numerically, the 
true potential of this variational method lies
in the relative ease in which approximations for
global flows may be solved,
and approximate solutions of LPAs found for
more than one invariant, which thus correspond to (possibly high 
dimensional\ref\ui{T.R. Morris, Phys. Lett. B357 (1995) 225.}) partial
differential equations. 

Should the $c$ function be extended beyond the LPA, 
we expect that the variational method 
will prove to be even more powerful.
\vfill\eject

\bigskip\medskip
\centerline{
\epsfxsize=0.8\hsize\epsfbox{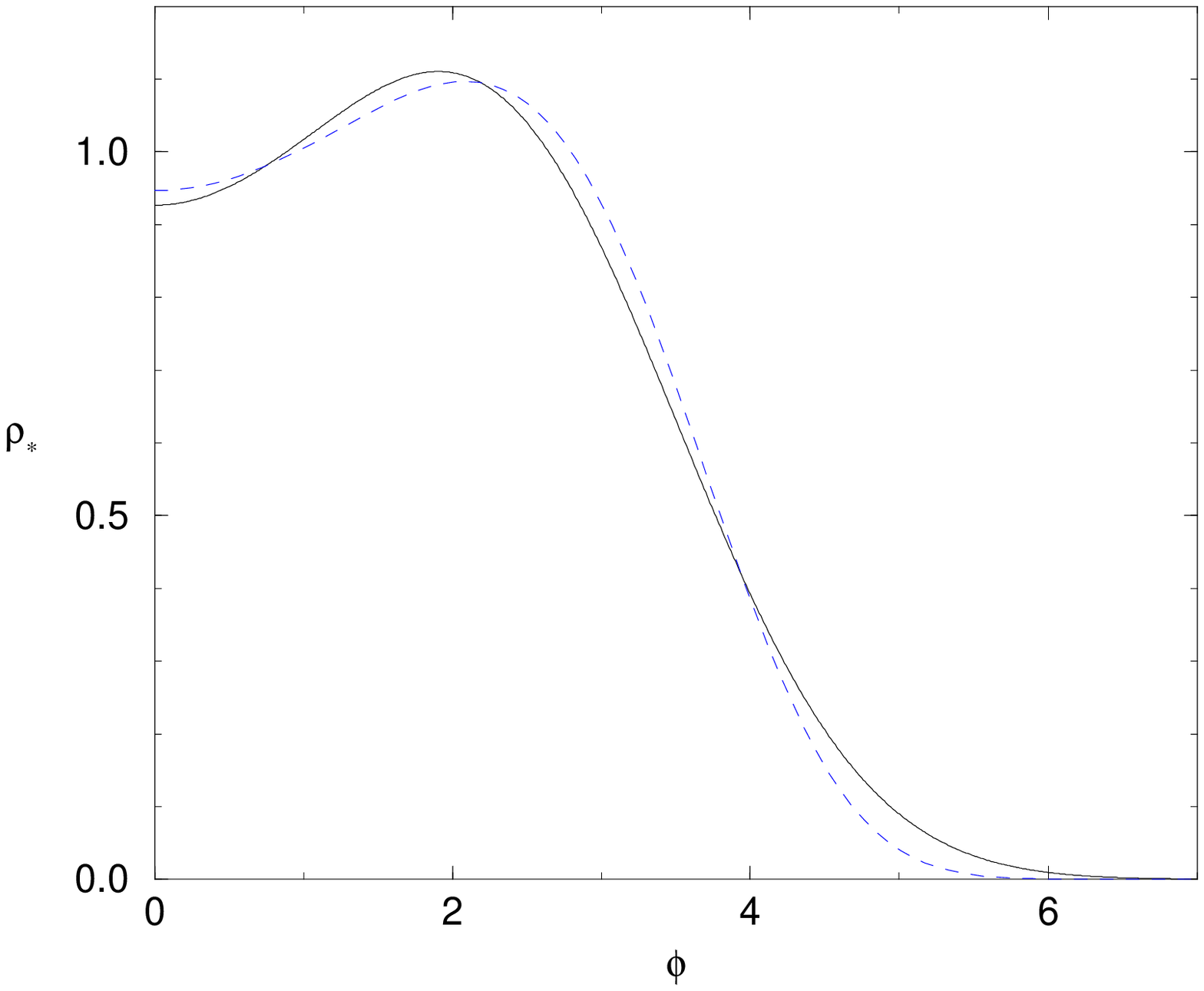}}  
\bigskip
\centerline{\vbox{\noindent {\bf Fig.1.} 
Plotted as $\rho_*$, the simplest polynomial variational approximation 
to the non-trivial fixed point potential (dashed line) is compared to the exact
solution to \lpa\ (full line).
}}
\bigskip\medskip

\bigbreak\bigskip\bigskip\bigskip\bigskip
\centerline{{\bf Acknowledgements}}\nobreak
TRM would like to thank John Cardy for helpful comments, and acknowledges
support of the SERC/PPARC through an Advanced Fellowship, and PPARC grant
GR/K55738. CHF and JMG acknowledge the support, through studentships, by
PPARC and the University of Southampton, respectively.

\listrefs

\end